\documentclass[prl,aps,twocolumn,amsmath,amssymb]{revtex4}
\usepackage{bm,graphicx,amsmath}
\usepackage{bbm}

\usepackage{bm}
\usepackage{amssymb}
\usepackage{epsfig}
\usepackage{epstopdf}
\usepackage{amstext}
\usepackage{latexsym}
\usepackage{times}

\begin{document}

\title{Cold-atom induced control of an opto-mechanical device}

\author{ M. Paternostro$^1$, G. De Chiara$^2$, and G. M. Palma$^3$}
\affiliation{$^1$School of Mathematics and Physics, Queen's University, Belfast BT7 1NN, United Kingdom\\
$^2$Grup de F{\'i}sica Te\`orica, Departament de F{\'i}sica, Universitat Aut{\`o}noma
de Barcelona, E-08193 Bellaterra, Spain\\
$^3$NEST-INFM (CNR) and Dipartimento di Scienze Fisiche ed Astronomiche, Universit\`a degli Studi di Palermo, Via Archirafi 36, I-90123 Palermo, Italy}

\begin{abstract}
We consider an optical cavity with a light vibrating end-mirror and containing a Bose-Einstein condensate (BEC). The mediation of the cavity field induces a non-trivial interplay between the mirror and the collective oscillations of the intra-cavity atomic density. We explore the thermodynamical implications of this dynamics and highlight the possibilities for indirect diagnostic. The effects we discuss can be observed in a set-up that is well within reach of current experimental capabilities and is central in the quest for mesoscopic quantumness.
\end{abstract}

\date{\today}

\maketitle

Achieving quantum control over a system endowed with macroscopic degrees of freedom is a long-sought goal of modern physics. The accomplishment of such a task will help us shifting the domain of applicability and exploitation of quantum technology from the context of microscopic systems fulfilling stringent criteria for quantumness to the mesoscopic world~\cite{arndt99,friedman2000,marshall2003}. Important fundamental and technological progresses have been performed, recently, along these directions: the gap separating current experimental possibilities from the observation of genuine quantum mechanical effects at the meso-scale is now only a few {\it quanta} wide~\cite{schwab2009A}. In such a quest, a few physical systems have emerged as well-suited for the observation of interesting quantum effects at a magnified scale: mechanical devices in optical and microwave resonators~\cite{natures2006A,groeblacher2009A}, collective excitations of ultracold atomic ensembles~\cite{esteve2008}
and arrays of superconducting devices~\cite{lehnert2008}. 

In this paper, we demonstrate mutual back-action dynamics of two macroscopic degrees of freedom embodied by physical systems of different nature. We consider the interplay between a BEC and the vibrating end-mirror of an optical cavity. We show a non-trivial intertwined dynamics between collective atomic modes, coupled to the cavity field, and the mechanical one, which experiences radiation-pressure forces.  By focusing on noise properties, important signatures of one subsystem in the dynamics of the other can be revealed by looking at experimentally accessible quantities. We characterize the atom-induced back-action that modifies the cooling capabilities of the opto-mechanical system and show that our predictions can be tested with current state of the art. Our study paves the way towards the use of the mutual interaction between atomic and mechanical subsystems for the sake of coherent quantum control at the mesoscopic scale. Such coupling has the potential for the implementation of effective schemes for quantum state engineering and dynamical manipulation of the (inaccessible) mechanical mode through the atomic subsystem.

\begin{figure}[t]
\psfig{figure=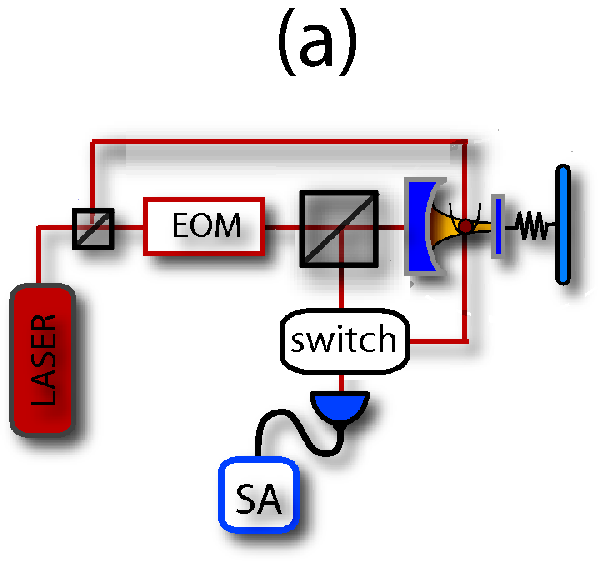,width=0.25\linewidth}
\psfig{figure=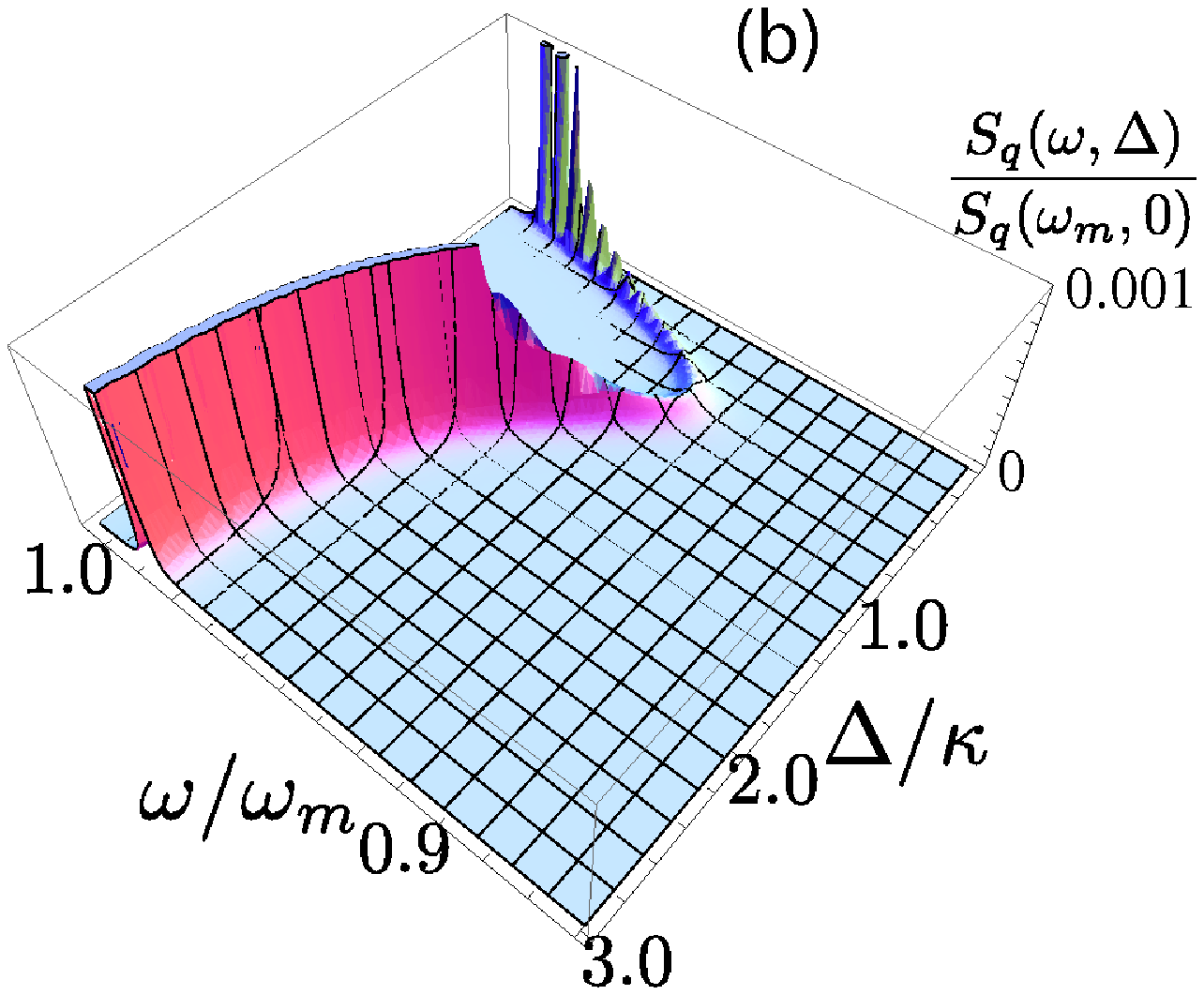,width=0.31\linewidth}
\psfig{figure=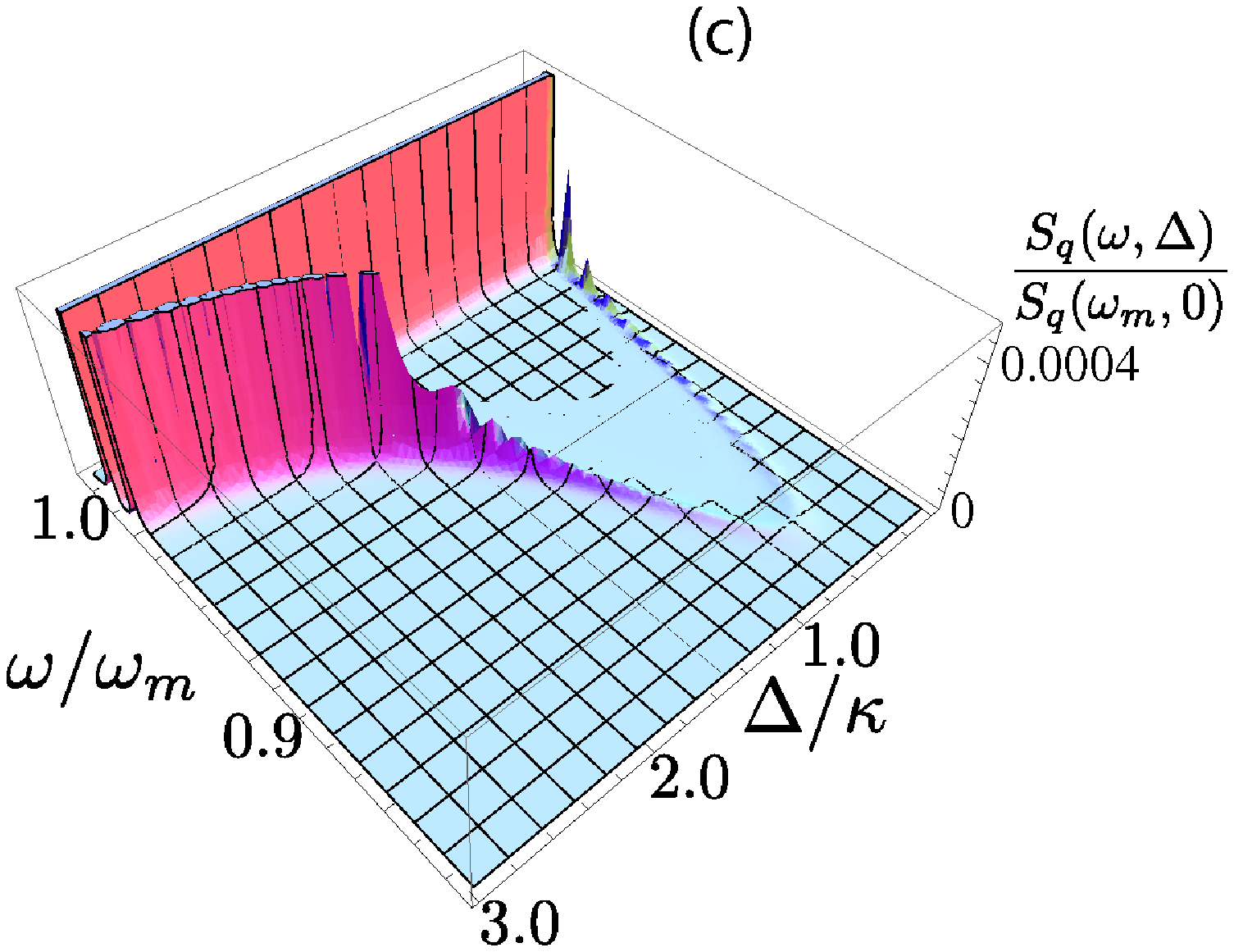,width=0.31\linewidth}
\caption{{\bf (a)} A laser is split by an umbalanced beam splitter. The transmitted part is phase-modulated at an electro-optics modulator (EOM) and enters the cavity coupled to a BEC. The (weak) reflected part of the pump laser probes the BEC. The signals from the cavity and the BEC go to a detection stage consisting of a switch (selecting the signal to analyze), a photodiode and a spectrum analyzer (SA). {\bf (b)} DNS $S_q(\omega,\Delta)$ for an empty cavity against $\Delta$ and $\omega$ for ${\cal L}{=}2.5$cm, $m{=}15$ng, $\omega_m/2\pi{=}{275}$KHz, $\gamma{=}\omega_m/Q$ with $Q{=}10^5$ and $T{=}300$K and $\kappa{\simeq}{5}$MHz.  The pumping light has wavelength $1064$nm and input power $4$mW. The DNS is rescaled to its value at $\omega{=}\omega_m$ and $\Delta{=}0$. {\bf (c)} We include the effects of the atomic coupling by taking $\tilde{\omega}{=}\omega_m$ and $\zeta{\simeq}0.7\chi{\sqrt{\hbar/(m\omega_m)}}$. 
}
\label{spettri}
\end{figure}

The movable end-mirror of the optical cavity of length ${\cal L}$ is assumed to perform harmonic oscillations at frequency $\omega_m$ along the cavity axis. The mirror is in contact with a background of  phononic modes in equilibrium at temperature $T$. The cavity is pumped through its (steady) input mirror by a laser of tunable frequency. The BEC is confined in a large-volume trap within the cavity~\cite{esteve2008,Esslinger2008} [cfr.~Fig.~\ref{spettri} {\bf (a)}]. Alternatively, the BEC could be sitting in a 1D optical-lattice generated by a trapping mode sustained by a bimodal cavity~\cite{Stamper2008}. The atom-cavity interaction is insensitive to the details of the trapping and our study holds in both cases. In the weakly interacting regime~\cite{StringariPitaevskii}, the atomic field operator can be split into a classical part (the condensate wave function) and a quantum one (the fluctuations) expressed in terms of Bogoliubov modes. Recent experiments coupling a BEC to an optical resonator~\cite{Esslinger2008} suggest that the Bogoliubov modes interacting significantly with the cavity field are those with momentum $\pm2k_c$ ($k_c$ is the cavity-mode momentum) while the condensate can be considered to be initially at zero temperature. The cavity end-mirror experiences radiation pressure while optical forces excite superpositions of atomic momentum modes. Interference between momentum-excited atoms and  condensate creates a periodic density grating sensed via the cavity. 

We write the Hamiltonian of the system made out of the cavity field, the movable mirror and the BEC as
$\hat{\cal H}{=}\sum_{j=M,C,A}\hat{\cal H}_j{+}\hat{\cal H}_{AC}{+}\hat{\cal H}_{MC}$
where the mirror, cavity and BEC Hamiltonians are 
${\hat{\cal H}_M=m \omega_m^2 \hat{q}^2/2 +{\hat{p}^2}/{(2m)}}$,
 ${\hat{\cal H}_C{=}\hbar (\omega_C{-}\omega_L)\hat{a}^\dagger\hat{a}{-}i \hbar \eta (\hat{a}-\hat{a}^\dagger)}$ and ${\hat{\cal H}_A{=}\hbar \tilde\omega\hat{c}^\dagger\hat{c}}$, respectively.
Here $\hat{q}$ ($\hat{p}$) is the mirror displacement (momentum), $m$ is its effective mass, $\omega_C$ ($\omega_L$) is the cavity (pump laser) frequency and $\hat{a}$ ($\hat{a}^\dag$) is the corresponding annihilation (creation) operator. Finally, $\tilde\omega$ and $\hat{c}$ ($\hat{c}^\dag$) are the frequency and the bosonic annihilation (creation) operator of the Bogoliubov mode. We have incorporated a {\it displacing} term $-i\hbar\eta(\hat{a}-\hat{a}^\dag)$ in the cavity Hamiltonian. This arises from the pump-cavity coupling, which shifts the cavity field in phase space (and, in turn, the equilibrium position of the vibrating end-mirror) proportionally to the coupling parameter $\eta{=}\sqrt{2\kappa{\cal R}/\hbar\omega_L}$ (${\cal R}$ is the laser power and $\kappa$ is the cavity decay rate). For small mirror displacements and large cavity free spectral range with respect to $\omega_m$ (so as to neglect scattering of photons into other mechanical modes), the mirror-cavity interaction can be put in the linearized form ${\hat{\cal H}_{MC}{=}-\hbar \chi \hat{q}\hat{a}^\dagger\hat{a}}$ with $\chi{=}\omega_C/{\cal L}$. On the other hand, the atoms-cavity interaction reads
\begin{equation}
\label{eq:HAC}
\hat{\cal H}_{AC}=({\hbar g^2 N_0})/({2\Delta_a})\hat{a}^\dagger\hat{a} + \hbar\sqrt{2}\zeta\hat{Q}\hat{a}^\dagger\hat{a},
\end{equation}
which contains two contributions: one is from the condensate only while the second is related to the position-like operator $\hat{Q}{=}(\hat{c}+\hat{c}^\dag)/\sqrt{2}$ of the Bogoliubov mode (we have assumed that the condensate wave function is not affected by the coupling to the cavity field). In Eq.~\eqref{eq:HAC}, $g$ is the vacuum Rabi frequency for the dipole-like transition connecting the atomic ground and excited states, $N_0$ is the number of condensed atoms, $\Delta_a$ is the detuning of the atomic transition from the cavity frequency and the coupling rate ${\zeta{\propto}\sqrt{N_0}{g^2}/\Delta_a}$. As discussed in~\cite{inpiu}, a rigorous calculation shows that $\zeta$ also depends on the Bogoliubov mode-function and can be conveniently tuned. While the first term in Eq.~\eqref{eq:HAC} embodies a cavity-frequency pull, the second is formally analogous to $\hat{\cal H}_{MC}$ and shows that, under the above working conditions, the BEC dynamics mimics that of a mechanical mode undergoing radiation-pressure effects. A similar result, for a BEC coupled to a static cavity, has been found in~\cite{Esslinger2008}. Our approach can be extended to include higher-order momentum modes in the expansion above. 

In general, the dynamics encompassed by $\hat{\cal H}$ is made difficult by the non-linearity inherent in $\hat{\cal H}_{MC}$ and Eq.~\eqref{eq:HAC}. However, for an intense pump laser, the problem can be linearized by introducing quantum fluctuations as $\hat{O}{\rightarrow}{\cal O}_s{+}\delta\hat{O}$ with $\hat{\cal O}$ any of the operators entering $\hat{\cal H}$, ${\cal O}_s$ its mean value and $\delta\hat{O}$  the associated first-order quantum fluctuation~\cite{Vitali2007}. We define the cavity quadratures $\hat{x}{=}\hat{a}{+}\hat{a}^\dagger$ and $\hat{y}{=}i(\hat{a}^\dagger{-}\hat{a})$ and the operator $\hat{P}{=}i(\hat{c}^\dag{-}\hat{c})/\sqrt{2}$ that is canonically conjugated to the position-like atomic operator $\hat{Q}$. The dynamical equations of the coupled three-mode system can then be cast into a compact form. Any realistic description of the problem at hand should include the most relevant sources of noise affecting the overall device, {i.e.} energy leakage from the cavity and thermal Brownian motion at temperature $T$ undergone by the cavity end-mirror. We thus consider the Langevin equation
$\partial_t{\hat{\bm \phi} }{=}{{\cal K}}\hat{\bm \phi}{+}\hat{\cal {\bm N}},$
where we have introduced the vector of fluctuations $\hat{\bm \phi}^T{=}(\delta\hat{x}~\delta\hat{y}~\delta\hat{q}~\delta\hat{p}~\delta\hat{Q}~\delta\hat{P})$,  the noise vector $\hat{\cal\bm N}^T{=}(\!\sqrt{\kappa}(\delta\hat{a}^\dag_{in}{+}\hat{a}_{in})~i\sqrt{\kappa}\delta(\hat{a}^\dag_{in}{-}\hat{a}_{in})~0~\hat{\xi}~0~0)$ and the dynamical coupling matrix ${\cal K}$, which is given in Ref.~\cite{inpiu}. The evolution of the system depends on a few crucial parameters, including the total detuning ${\Delta{=} \omega_C{-}\omega_L{-}\chi q_s{+}\sqrt 2\zeta Q_s{+}g^2N_0/2\Delta_a}$ between the cavity and the pump laser. This consists of the steady pull-off term in Eq.~\eqref{eq:HAC} as well as both the opto-mechanical contributions proportional to the displaced equilibrium positions  of the mechanical and Bogoliubov modes. These are respectively given by the stationary values ${q_s{=}{\hbar \chi\alpha_s^2}/{m\omega_m^2}}$ and ${Q_s{=-}{\sqrt 2 \zeta\alpha_s^2}/{\tilde\omega}}$, which are in turn determined by the mean intra-cavity field amplitude $\alpha_s{=}{\eta}/\sqrt{{\Delta^2+\kappa^2}}$. The interlaced nature of such stationary parameters (notice the dependence of $\alpha_s$ on the detuning) is at the origin of bistability and chaotic effects~\cite{Esslinger2008,Stamper2008,Meystre2009}. As for the noise-related part of the dynamics, we have introduced  $\delta\hat{a}_{in}$ and $\delta\hat{a}^\dag_{in}$ as zero-average [${\langle a_{in}(t)\rangle{=}\langle a^\dag_{in}(t)\rangle{=}0}$], delta-correlated [$\langle a_{in}(t) a_{in}^\dagger(t') \rangle{=}\delta(t{-}t')$] operators describing white noise entering the cavity from the leaky mirror. Dissipation of the mechanical mirror energy is, on the other hand, associated with the decay rate $\gamma$ and the corresponding zero-mean Langevin-force operator $\hat{\xi}(t)$ having non-Markovian correlations ($\beta_B{=}\hbar/2k_BT$)~\cite{Giovannetti2001} $\langle\hat{\xi}(t)\hat{\xi}(t') \rangle=({\hbar\gamma m}/{2\pi})\int\!\omega e^{-i \omega(t-t')}[ \coth({\beta_B\omega}){+}1] d\omega$. Although the non-Markovianity of the mechanical Brownian motion could be retained in our approach, for large mechanical quality factors ($\gamma{\rightarrow}{0}$), a condition that is met in current experiments on micro-mechanical systems~\cite{groeblacher2009A}, one can take ${\langle\hat{\xi}(t)\hat{\xi}(t')\rangle{\simeq}[\hbar\gamma{m}/\beta_B{+}i\partial_t]\delta(t{-}t')}$~\cite{Giovannetti2001}. As our analysis relies on symmetrized two-time correlators, the antisymmetric part in the above expression, proportional to $\partial_t\delta(t{-}t')$, is ineffective, thus making our description fully Markovian. Here we show that the model above results in an interesting back-action dynamics where the state of the mechanical mode is strongly intertwined with the BEC. The physical properties of the mirror are altered by the cavity-BEC coupling. Evidences of such interaction, strong enough to inhibit the cooling capabilities of the radiation-pressure mechanism under scrutiny, are found in the noise properties of the mechanical mode.  

We start considering the modification in the mirror dynamics due to the coupling to the cavity and indirectly to the BEC. The Langevin equations are solved in the frequency domain, where we should ensure stability of the solutions. This implies 
negativity of the real part of the eigenvalues of ${\cal K}$. 
Numerically, we have fount that stability is given for ${\Delta{>}0}$ and weak coupling of the mirror and the BEC to the cavity, {i.e.} for $\{{\chi\sqrt{\hbar/m\omega_m}, \zeta\}{\ll}\kappa}$, which are conditions fulfilled throughout this work. 
We find the mirror displacement
$\delta\hat{q}(\omega){=}[{\cal A}_M(\omega)\delta\hat{y}_{in}(\omega) + {\cal B}_M(\omega)\delta\hat{x}_{in}(\omega)+{\cal C}_M(\omega)\hat{\xi}(\omega)]$,
with ${\cal A}_M(\omega){=}{\cal B}(\omega)\Delta/(\kappa{-}i\omega){=-}\hbar\chi\alpha_s{\sqrt{2\kappa}}\Delta/d_M(\omega)$,
${\cal C}_M(\omega){=-}\{(\omega^2{-}\tilde\omega^2)[(\kappa{-}i\omega)^2{+}\Delta^2]{+}4\tilde\omega\Delta\alpha_s^2\zeta^2\}/d_M(\omega)$
and $d_M(\omega)$ that is related to the effective susceptibility function of the mechanical mode~\cite{inpiu}. 
We now compute the density noise spectrum (DNS) of $\delta\hat{q}(\omega)$. For a generic operator $\hat{O}(\omega)$ in the frequency domain, the DNS is defined as ${S\!_{\cal O}(\omega)\!=\!({1}/{4\pi})\int d\Omega e^{-i(\omega+\Omega)t} \langle\hat{\cal O}(\omega)\hat{\cal O}(\Omega){+}\hat{\cal O}(\Omega)\hat{\cal O}(\omega)\rangle}$.
Using the correlation properties of the input and Brownian noise operators, after a little algebra one gets 
\begin{equation}
\label{spettroM}
S\!_q(\omega){=}\sum_{{\cal J}={\cal A},{\cal B}}|{\cal J}_M(\omega)|^2+\hbar\gamma m\left[1\!+\!\coth({\beta\omega})\right]|{\cal C}_M(\omega)|^2.
\end{equation}
Some interesting features emerge from the study of $S\!_q(\omega)$. In Fig.~\ref{spettri} we compare  the case of an empty opto-mechanical cavity [panel {\bf (b)}] and one where a weak coupling with the atomic Bogoliubov mode of frequency ${\tilde{\omega}{=}\omega_{\text{m}}}$ is  included [panel {\bf (c)}]. For an empty cavity, the mechanical-mode spectrum is obviously identical to what has been found in Ref.~\cite{MauroNJP} (the use of that case as a milestone in our quantitative study motivates the choice of the parameters used throughout this work). Both the optical spring effect in a detuned optical cavity and a cooling/heating mechanism are evident: height, width and peak-frequency of $S\!_q(\omega)$ change with the detuning $\Delta$. At ${\Delta\simeq{\kappa/2}}$ optimal cooling is achieved with a considerable shrink in the height of the spectrum. However, as soon as the Bogoliubov mode enters the dynamics, major modifications appear. The optical spring effect is magnified (the red-shift of the peak frequency of $S\!_q(\omega)$ is larger than at $\zeta{=}0$) and a secondary structure appears in the spectrum, unaffected by any change of $\Delta$. Such a structure is a second Lorentzian peak centered in ${\omega{\simeq}\omega_{\text{m}}}$ and is a signature of the back-action induced by the atoms, an effect that comes from a three-mode coupling and, as discussed later, is {\it determined} by $\tilde{\omega}$ and $\zeta$. In fact, by studying the dependence of $S\!_q(\omega)$ on the frequency of the Bogoliubov mode, we see that the secondary peak identified above is centered at $\tilde{\omega}$. For ${\zeta\ll{\chi}\sqrt{\hbar/(m\omega_m)}}$, {\it i.e.} for weak back-action from the atomic mode onto the mechanical one, the signature of the former in the spectrum of the latter is small. A quantitative assessment reveals, in fact, that it only consists of a tiny structure subjected to negligible detuning-induced changes. The picture changes for $\tilde\omega$ close to the mechanical frequency. In this case, as seen in Fig.~\ref{spettri} {\bf (c)}, the influence of the atomic medium is considerable and present at any value of $\Delta$. While the mechanical mode experiences enhanced optical spring effect (as easily seen by looking at the effective susceptibility of the mechanical mode~\cite{inpiu}), the secondary structure persists even at ${\Delta{\sim}{\kappa}/2}$, the working point that for our choice of parameters optimizes the mechanical cooling at empty cavity. However, as demonstrated later on, here the strong optical spring effect is not accompanied by an effective mechanical cooling. 

\begin{figure}[b]
{\bf (a)}\hskip3cm{\bf (b)}\hskip3cm{\bf (c)}\\
\psfig{figure=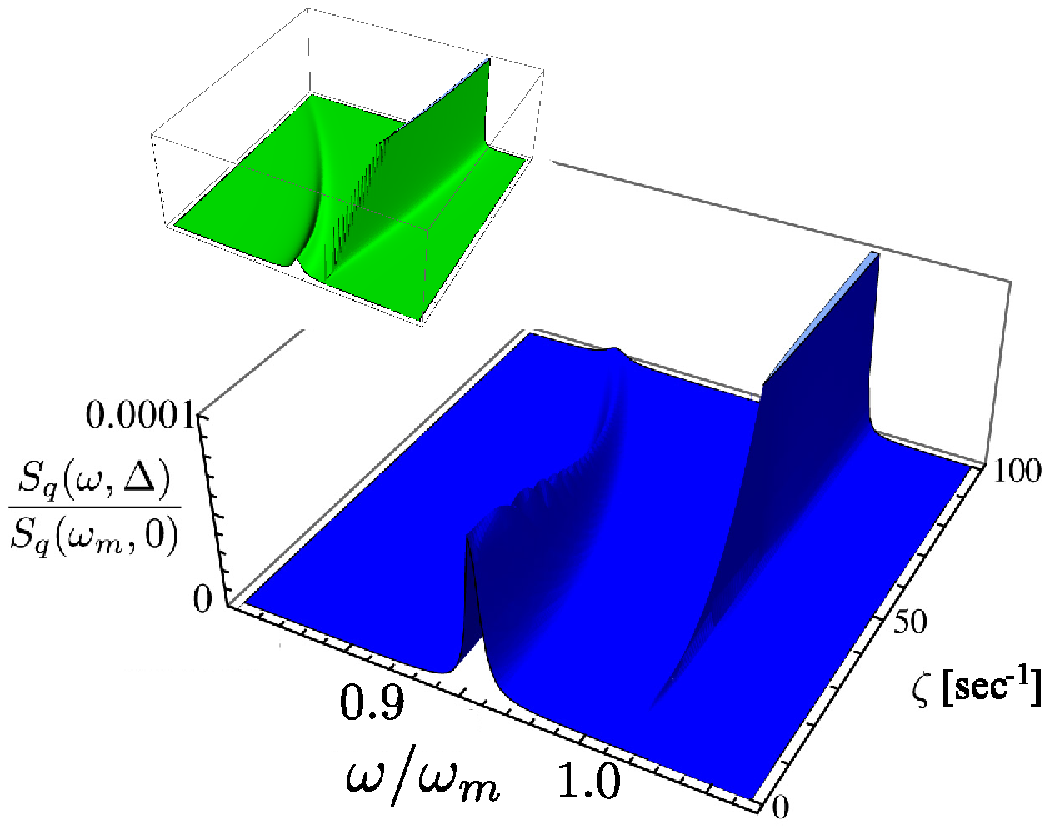,width=0.331\linewidth}
\psfig{figure=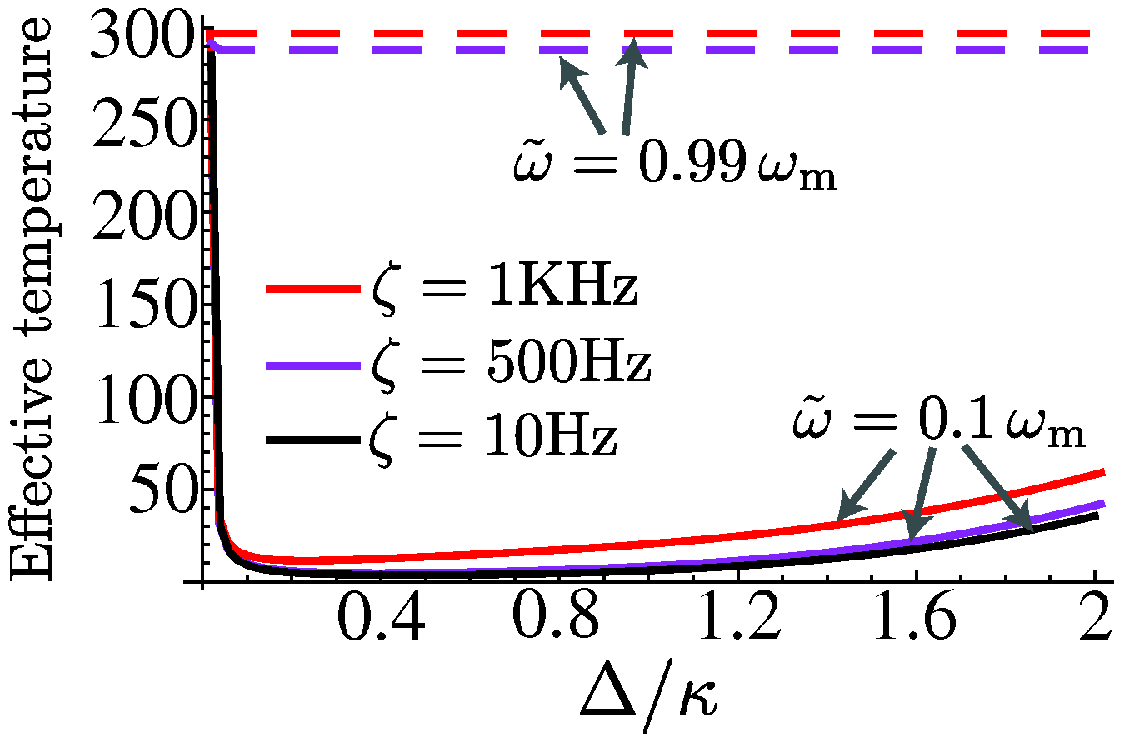,width=.30\linewidth}
\psfig{figure=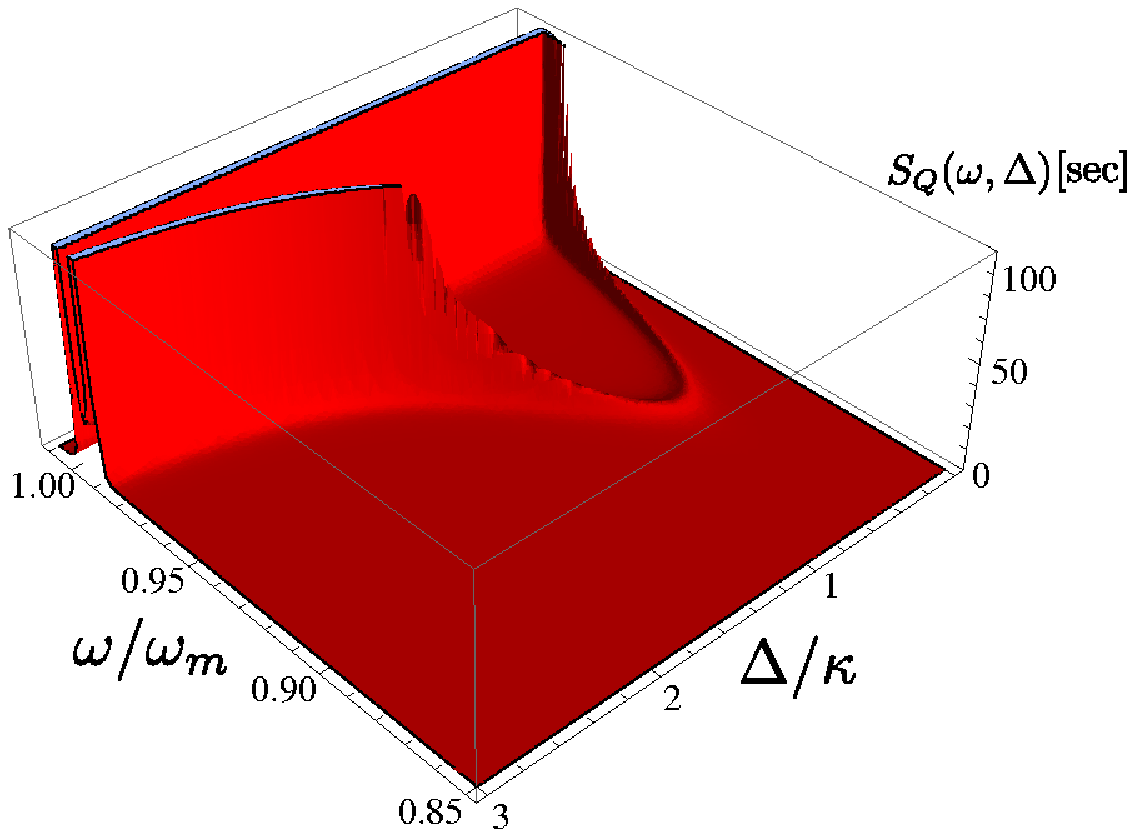,width=.33\linewidth}
\caption{{\bf (a)} DNS of the mechanical mode against $\omega$ and $\zeta$ for ${\tilde{\omega}{\simeq}{\omega_m}}$ and ${\Delta{=}\kappa/2}$. The structure centered at $\omega_m$ and with amplitude growing with $\zeta$ is due to atomic back-action. 
Inset: same plot for $\tilde{\omega}{=}0.8\omega_m$. Similar but less important features are found away from the resonance betwen mechanical and atomic mode.  {\bf (b)} Temperature of the mechanical mode against $\Delta/\kappa$. Solid (Dashed) lines are for $\tilde{\omega}{=}0.1\omega_{\text{m}}$ ($\tilde{\omega}{\simeq}\omega_{\text{m}}$). 
{\bf (c)} $S\!_Q(\omega,\Delta)$ for $\tilde\omega{\simeq}{\omega}_m$ and $\zeta{=}50$Hz. }
\label{backaction}
\end{figure}

A better understanding is provided by studying $S\!_q(\omega,\Delta)$ against the atomic opto-mechanical rate $\zeta$ [cfr. Fig.~\ref{backaction} {\bf (a)}]. At the optimal empty-cavity detuning and for $\tilde{\omega}{\simeq}{\omega}_m$, both the effects highlighted above are clearly seen: the {\it contribution} of the secondary structure centered at $\tilde{\omega}$ grows with $\zeta$ due to the increasing atomic back-action while a large red-shift and shrinking of the mechanical-mode contribution to the DNS shows the enhanced spring effect. An intuitive explanation for all this comes from taking a normal-mode description, where the diagonalization of $\hat{\cal H}$ passes through the introduction of new modes that are linear combinations of the mechanical and Bogoliubov one. The weight of the latter increases with $\zeta$, thus determining a strong influence of the atomic part of the system over the noise properties of the mechanical mode.  

The consequences of the atomic back-action are not restricted to the effects highlighted above. Strikingly, the coupling between the atomic medium and the cavity field acts as a {\it switch} for the cooling experienced by the mechanical mode in an empty cavity  [cfr.~Fig.~\ref{backaction} {\bf (b)}]. That is, the coupling to the collective oscillations of the atomic density is crucial in determining the number of thermal excitations in the state of the mechanical mode, regulating the mean energy of the cavity end-mirror. A way to clearly see it is to consider the effective temperature $T_{\text{eff}}{=}\langle U\rangle/k_B$, where ${\langle U\rangle{=}m\omega^2_m\langle\delta\hat{q}^2\rangle/2{+}\langle\delta\hat{p}^2\rangle/(2m)}$ is the mean energy of the mechanical mode. $\langle U\rangle$ is experimentally easily determined by measuring just the area underneath $S\!_q(\omega,\Delta)$, as acquired by a spectrum analyzer. In fact, we have ${\langle\delta\hat{r}^2\rangle{=}\int{d}\omega{S}\!_r(\omega,\Delta)~(r{=}q,p)}$ with ${S\!_p(\omega,\Delta){=}m^2\omega^2_mS\!_q(\omega,\Delta)}$. 
Such temperature-regulating mechanism is explained in terms of a simple thermodynamic argument. The exchange of excitations behind passive mechanical cooling~\cite{natures2006A,MauroNJP} occurs at the optical sideband centered at $\omega_m$. When the frequency of the Bogoliubov mode does not match this sideband, mirror and cavity field interact with only minimum disturbance from the BEC. Thus, mechanical cooling occurs as in an empty cavity: even for relatively large values of $\zeta$ the cooling capabilities of the detuned opto-mechanical process are, for all practical purposes, unaffected [see Fig.~\ref{backaction} {\bf (b)}]. However, by tuning $\tilde\omega$ on resonance with the relevant optical sideband, we introduce a {\it well-source mechanism} for the recycling of phonons extracted from the mechanical mode and transferred to the cavity field. The BEC can now {absorb} some excitations taken from the mirror by the field, thus acting as a {\it phononic  well} and {release} them into the field at a frequency matched with $\omega_m$. The mirror can take the excitations back, as in the presence of a {\it phononic source}: thermodynamical equilibrium is established at a temperature set by $\zeta$. For strong atomic back-action, the mirror does not experience any cooling [Fig.~\ref{backaction} {\bf (b)}].

\begin{figure}[t]
{\bf (a)}\hskip3.cm{\bf (b)}\hskip3.cm{\bf (c)}\\
\psfig{figure=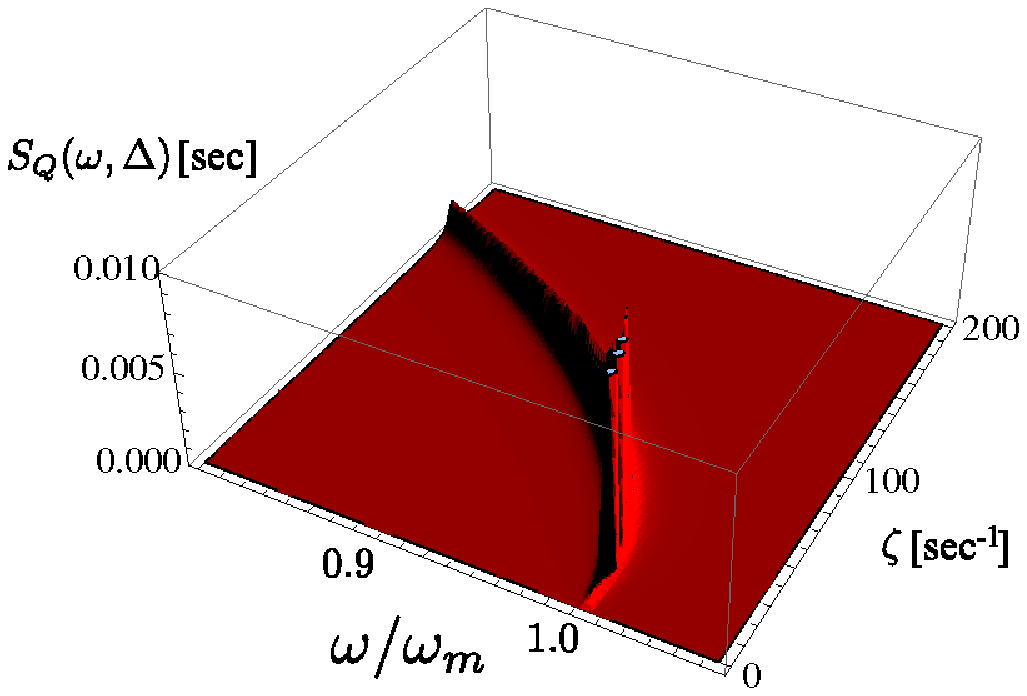,width=.32\linewidth}
\psfig{figure=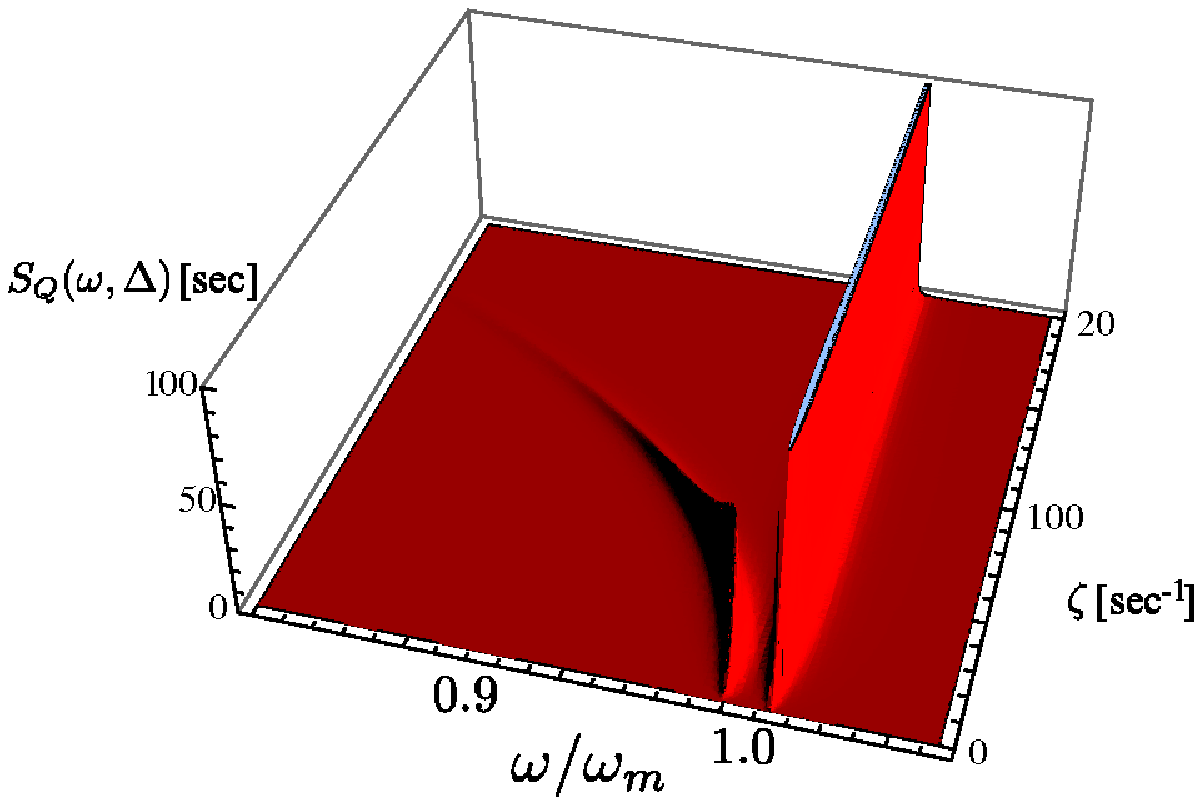,width=.32\linewidth}
\psfig{figure=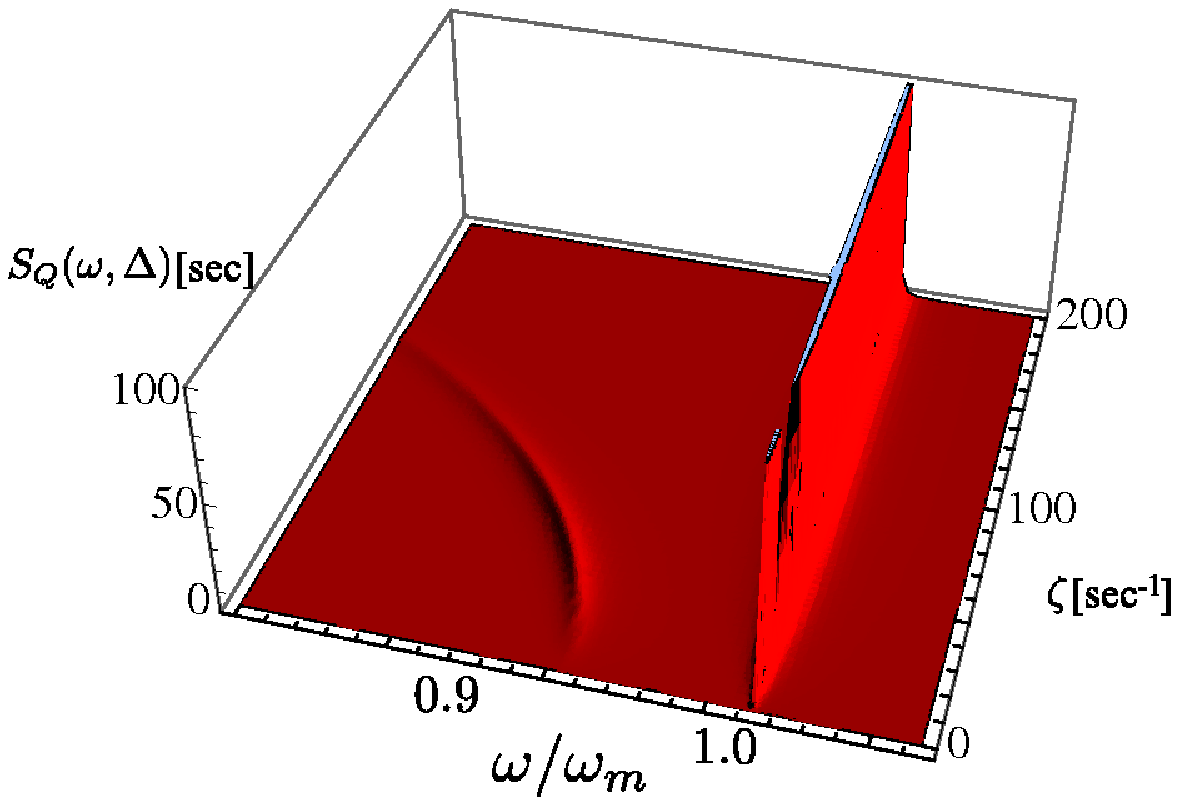,width=.32\linewidth}
\caption{Panels {\bf (a)} to {\bf (c)}: $S\!_Q(\omega,\kappa/2)$ against $\omega$ and $\zeta$ for $\chi{=}k\omega_C/2{\cal L}~(k{=}0,1,2)$. The atomic and mechanical part of the spectrum are clearly splitted.}
\label{multiplatomi}
\end{figure}

Analogously, one finds the atomic DNS associated with the position-like operator of the Bogoliubov mode, which reads $\delta\hat{Q}(\omega)=[{\cal A}_A(\omega)\delta\hat{y}_{in}+{\cal B}_A(\omega)\delta\hat{x}_{in}+{\cal C}_A(\omega)\hat{\xi}(\omega)]$ with ${\cal A}_A(\omega){=}\Delta{\cal B}(\omega)/[\tilde{\omega}(\kappa{-}i\omega)]{=}2i\alpha_s\zeta\Delta\sqrt{\kappa}\omega(i\gamma\omega{+}\omega^2{-}\omega^2_m)/d_{A}$, ${{\cal C}_A(\omega)=-2\sqrt{2}i\alpha^2_s\Delta\zeta\chi\omega\tilde{\omega}/md_{A}}$ and $d_A$ being rather lengthy. The spectrum ${S}\!_{Q}(\omega,\Delta)$ is then easily determined using the appropriate input-noise correlation functions and sketched in Fig.~\ref{multiplatomi}. Clearly, in light of the formal equivalence of Eq.~\eqref{eq:HAC} with a radiation pressure mechanism, by setting up the proper working point, the BEC should undergo a cooling dynamics similar to the one experienced by the mirror. The starting temperature of the Bogoliubov mode depends on the values taken by $\tilde{\omega}$ and $\zeta$. At $\zeta{=}0$, regardless of the atomic-mode frequency, its effective temperature is very low, as it should be. For a set value of $\zeta$, the temperature arises as $\tilde{\omega}{\rightarrow}\omega_m$. 
The conditions of our investigations are such that weak coupling between the BEC and the cavity field are kept, in a way so as to make the Bogoliubov expansion valid and rigorous. 
The mutually-induced back-action at the center of our discussion is clearly visible in Figs.~\ref{backaction} {\bf (c)} and \ref{multiplatomi}, where features similar to those present in the mechanical DNS appear. For ${\chi{=}0}$, the atomics DNS at $\tilde{\omega}{=}\omega_m$ starts from zero (at ${\zeta{=}0}$) and experiences red shifts and shrinking as the effective opto-mechanical coupling rate grows. Having switched off the coupling between the mechanical mode and the field, the spectrum is single-peaked. This is not the case for ${\chi{\neq}{0}}$ where a secondary structure appears, similar to the one in the mechanical DNS. The splitting between mechanical and atomic contributions to $S\!_Q(\omega,\kappa/2)$ grows with $\chi$, a sign of the mechanically-enhanced effect felt by the atomic mode.

We have demonstrated an interesting mechanism of dynamical back-action within an experimentally viable set-up. The indirect mutual cross-talking between mechanical and atomic mode determines a substantial modification of the respective dynamics and leaves a signature of the reciprocal influence in experimentally handy quantities. This opens up the way to novel diagnostic strategies where the relevant interaction parameters are determined by measuring the noise properties of only one subsystem and fitting it with the analytical expressions for the relevant spectra provided here. The BEC spectrum could be probed by homodyning a weak forward-scattered field, transversally fed into the cavity and coupled to the atoms, as done in similar contexts~\cite{esteve2008} [cfr.~Fig.~\ref{spettri} {\bf (a)}]. The procedure for reconstructing the intra-cavity dynamics from measurements on the extra-cavity field described in~\cite{MauroNJP} could be adopted too. Our study demonstrates coherence and mutual control in the coupling between the BEC and the mirror. These features are crucial in designing strategies for state-engineering of the mechanical mode by means of its interaction with the BEC, which is more accessible, easier to manipulate and less prone to noise effects. More crucially, a protocol for {\it quantitative} entanglement-diagnostic in the mirror-field state performed through direct determination of the quantum correlations between BEC and field can be designed and it will be reported elsewhere~\cite{tocome}. This will open up the possibility for the direct and exact quantification of opto-mechanical entanglement without relying on demanding all-optical approaches~\cite{Vitali2007}.

{\it Acknowledgments.--} We thank T. Esslinger and J. F. McCann for comments. We acknowledge support from the Spanish Ministry of Science and Innovation through the program Juan de la Cierva and the projects FIS2008-01236 and QOIT (Consolider Ingenio 2010), EUROTECH and EPSRC (EP/G004579/1). MP and GMP thank the CQT, National University of Singapore, where part of the work has been done.


\begin{thebibliography}{99}

\bibitem{arndt99} M. Arndt, {\it et al.}, {\it Nature} (London) {\bf 401}, 680 (1999).

\bibitem{friedman2000} J. R. Friedman, {\it et al.}, {\it Nature} (London) {\bf 406}, 43 (2000).

\bibitem{marshall2003} W. Marshall, {\it et al.}, {\it Phys. Rev. Lett.} {\bf 91}, 130401 (2003).

\bibitem{schwab2009A} T. Rocheleau, {\it et al.}, {\it Nature} (London), {\bf 463}, 72 (2009); J. B. Hertzberg, {\it et al.}, {\it Nature Phys.}, (to appear 2009).

\bibitem{natures2006A} S. Gigan, {\it et al.}, {\it Nature} (London) {\bf 444}, 67 (2006); O. Arcizet, {\it et al.}, {\it ibid.} {\bf 444}, 71 (2006); A. Naik, {\it et al.}, {\it ibid.} {\bf 443}, 193 (2006).

\bibitem{groeblacher2009A} S. Gr\"oblacher, {\it et al.}, {\it Nature} (London) {\bf 460}, 724 (2009); J. D. Thompson, {\it et al.}, {\it ibid.} {\bf 452}, 72 (2008); S. Gr\"oblacher, {\it et al.}, {\it Nature Phys.} {\bf 5}, 485 (2009); A. Schliesser, {\it et al.},
{\it ibid.} {\bf 5}, 509 (2009).

\bibitem{esteve2008} J. Est\`eve, {\it et al.}, {\it Nature} (London) {\bf 455}, 1216 (2008).
J. F. Sherson, {\it et al.}, {\it ibid.} {\bf 443}, 557 (2006);
M. Greiner, {\it et al.}, {\it ibid.} {\bf 415}, 39 (2002);
F. Brennecke, {\it et al.}, {\it ibid.} {\bf 450}, 268 (2007); Y. Colombe, {\it el al.}, {\it ibid.} {\bf 450}, 272 (2007).

\bibitem{lehnert2008} C. A. Regal, J. D. Teufel, K. W. Lehnert, {\it Nature Phys.} {\bf 4}, 555 (2008); J. D. Teufel, {\it et al.}, {\it ibid.} {\bf 4}, 820 (2009).

\bibitem{StringariPitaevskii}
L. Pitaevskii and S. Stringari, {\it Bose-Einstein condensation} (Oxford University Press, Oxford, 2003).

\bibitem{Esslinger2008}
F. Brennecke, {\it et al.}, {\it Science} {\bf 322}, 235 (2008).

\bibitem{Stamper2008} K. W. Murch, {\it et al.}, {\it Nature Phys.} {\bf 4}, 561 (2008).


\bibitem{Meystre2009} K. Zhang, {\it et al.}, Phys. Rev. A {\bf 81}, 013802 (2010). 


\bibitem{Giovannetti2001} V.~Giovannetti~and~D.~Vitali,~{\it Phys.~Rev.~A}~{\bf 63}, 023812 (2001).

\bibitem{MauroNJP} M. Paternostro, {\it et al.}, {\it New J. Phys.} {\bf 8}, 107 (2006). 


\bibitem{Vitali2007} D. Vitali, {\it et al.},~{\it Phys.~Rev.~Lett.} {\bf 98}, 030405 (2007).


\bibitem{inpiu} See supplementary material at\\ http://link.aps.org/supplemental/10.1103/PhysRevLett.104.243602 for an additional
analysis of the properties of the system.

\bibitem{tocome} G. De Chiara, M. Paternostro, and G. M. Palma (to appear, 2010).

\end{thebibliography}
\end{document}